\pdfoutput=1
\documentclass{article}

\usepackage{arxiv}

\usepackage[utf8]{inputenc} 
\usepackage[T1]{fontenc}    
\usepackage{hyperref}       
\usepackage{url}            
\usepackage{booktabs}       
\usepackage{amsfonts}       
\usepackage{nicefrac}       
\usepackage{microtype}      
\usepackage{lipsum}
\usepackage{multirow}
\usepackage{amsmath}
\usepackage{graphicx}
\usepackage{float}
\usepackage{caption}
\floatstyle{plaintop}
\restylefloat{table}
\usepackage{multicol}
\usepackage{makecell}
\usepackage{comment}

\title{Study of Different Deep Learning Approach with Explainable AI for Screening Patients with \emph{COVID-19} Symptoms: Using CT Scan and Chest X-ray Image Dataset}

\author{Md Manjurul Ahsan \\
  University of Oklahoma\\
  Norman, USA\\
  \texttt{ahsan@ou.edu} \\
   \And
 Kishor Datta Gupta \\
  University of Memphis\\
  Memphis, Tennesse \\
  \texttt{kgupta1@memphis.edu}\\
  \and
 Mohammad Maminur Islam\\
  University of Memphis\\
  Memphis, Tennesse\\
  \texttt{mislam3@memphis.edu}\\
  \and
  Sajib Sen\\
  University of Memphis\\
  Memphis, Tennesse \\
  \texttt{ssen4@memphis.edu}\\
  \and
  Md. Lutfar Rahman\\
  University of Memphis\\
  Memphis, Tennesse \\
  \texttt{mrahman9@memphis.edu}\\
  \and
  Mohammad Shakhawat Hossain\\
  Marquette University\\
  Milwaukee, Wisconsin \\
  \texttt{mohammadshakhawat.hossain@marquette.edu}
  }

\begin{document}
\maketitle

\begin{abstract}
The outbreak of COVID-19 disease caused more than 100,000 deaths so far in the USA alone.  It is necessary to conduct an initial screening of patients with the symptoms of COVID-19 disease to control the spread of the disease. However, it is becoming laborious to conduct the tests with the available testing kits due to the growing number of patients.  Some studies proposed CT scan or chest X-ray images as an alternative solution. Therefore,  it is essential to use every available resource, instead of either a CT scan or chest X-ray to conduct a large number of tests simultaneously. As a result, this study aims to develop a deep learning-based model that can detect COVID-19 patients with better accuracy both on CT scan and chest X-ray image dataset. In this work, eight different deep learning approaches such as VGG16, InceptionResNetV2, ResNet50, DenseNet201, VGG19, MobilenetV2, NasNetMobile, and ResNet15V2 have been modified and tested on two dataset-one dataset includes 400 CT scan images, and another dataset includes 400 chest X-ray images. Result show that, NasNetMobile outperformed all other models in terms of accuracy on CT scan (81.5\% -- 95.2\%) and chest X-ray (95.4\% -- 100\%) image dataset with 95\% confidence interval.
Besides, Local Interpretable Model-agnostic Explanations (LIME) is used to explain the model's interpretability. Using LIME, test results demonstrate that it is conceivable to interpret top features that should have worked to build a trust AI framework to distinguish between patients with COVID-19 symptoms with other patients.
\end{abstract}


\section{Introduction}
The novel coronavirus, also known as COVID-19, created colossal health crises in 2020 worldwide. The virus that caused this disease known as severe acute respiratory syndrome coronavirus 2, also called SARS-CoV-2~\cite{stoecklin2020first}. Since the virus is spreading very fast, thus the number of infected people is increasing day by day, while the test kit is limited along with limited hospitals. While many COVID-19 cases exhibit mild symptoms, a small percentage suffers from severe or critical conditions~\cite{mckeever2020here}. In increasingly genuine cases, the contamination can cause pneumonia, extreme intense respiratory condition, multi-organ failure, and death~\cite{mahase2020coronavirus}. Developed countries like the USA, UK, and Italy, the health systems have been overwhelmed due to the expanding demand for intensive care units, as those units filled with COVID-19 patients with severe medical conditions \cite{tanne2020covid}.\\
Since COVID-19 is a socially transmitted disease,  screening patients with COVID-19 symptoms is the first and foremost step. As per the most recent rules acknowledged by the Chinese government, the determination of COVID-19 ought to be affirmed by gene sequencing for respiratory or blood tests as a key marker for reverse transcription-polymerase chain reaction (RT-PCR)~\cite{narin2020automatic}. However, while patients are waiting for the test results, many more people are affected by them, before moving them to isolation. Therefore, the earlier it is possible to detect COVID-19 patients, the faster the patients' isolation can be done to reduce community spreading of the disease.\\
Factors such as airspace opacities, ground-glass opacity(GGO), and later consolidation are important signs in the lung, which play an essential role in detecting COVID-19 patients \cite{wong2020frequency}. Due to the limitations of the test kit, several studies proposed alternative solutions like CT scan or chest X-ray images for early detection of COVID-19 patients~\cite{gozes2020rapid,Biodifferences2020diff,narin2020automatic,butt2020deep}.
 X-ray machines are used to examine the affected body, such as cracks, bone disengagements, lung contamination, pneumonia, and tumors. On the other hand, CT scanning is a  cutting edge X-ray machine that looks at the extremely delicate structure of the dynamic body part and more clear pictures of the delicate inward tissues and organs~\cite{Biodifferences2020diff}. Utilizing X-rays is a quicker, simpler, less expensive, and less perilous strategy than CT~\cite{narin2020automatic}. Pneumonia is one of the most significant indications of COVID-19 disease, and it opens holes in the lungs like SARS, giving them a "honeycomb-like appearance." Both CT scan or chest X-ray uses to diagnose pneumonia. Thus, chest X-ray or CT scans could be the best choice as an early screening method~\cite{narin2020automatic}.\\
\cite{shan2020lung} developed a deep learning-based model for segmenting infectious sites on the lung using chest CT. Butt\& Gill (2020), developed early detection techniques to distinguish between COVID-19 pneumonia and influenza patients using deep-learning techniques~\cite{butt2020deep}.~\cite{wang2020covid} also used deep learning techniques to extract features from CT images related to COVID-19 disease.\\
Similarly, some of the studies use chest X-ray to detect MERS-CoV and SARS-CoV, known as cousins of COVID-19~\cite{hamimi2016mers}. A study conducted by Kapur (2008) used data mining techniques to distinguish between pneumonia and SARS based on X-ray images~\cite{babaoglu2005lecture}.~\cite{narin2020automatic} proposed three different convolutional neural network-based models (ResNet50, InceptionV3 and InceptionResNetV2) to identify coronavirus pneumonia infected patients based on chest X-ray images.\\
To extend existing work one step further, in this experiment, eight different deep learning models (VGG16~\cite{simonyan2014very}, InceptionResNetV2~\cite{langkvist2014review}, ResNet50~\cite{akiba2017extremely}, DenseNet201~\cite{huang2017densely}, VGG19~\cite{simonyan2014very}, MobileNetV2~\cite{sandler2018mobilenetv2}, NasNetMobile~\cite{da2018lung} ,and ResNet15V2~\cite{varatharasan2019improving}) have been proposed for the detection of COVID-19 patients using CT scan and chest X-ray images. This research's novelty is summarized as follows: 1) Eight different convolutional neural network-based models have been proposed and tested on CT scan image and chest X-ray image dataset. 2) A comparative analysis was done in terms of accuracy, precision, recall, and f1-score. 3) Results showed MobileNetV2 and NasNetMobile outperformed all other models both on CT scan and chest X-ray image datasets. 4) Finally, models were explained with the help of Local Interpretable Model-agnostic Explanations (LIME).
\section{Literature Review}
\subsection{X-ray based Screening of COVID-19}
Several studies considered X-ray images due to its less sensitivity compared to CT scan images. A recent study showed that COVID-19 patients' early or mild symptoms are possible to detect using X-ray images~\cite{wong2020frequency}. Among them, 69\% of the patients' X-ray report showed abnormality at the initial time of admission, while 80\% of the patients' symptoms showed after their hospitalization.
Since the disease outbreak in 2020 worldwide, there was not enough data to study extensively during that time. For example, ~\cite{ghoshal2020estimating} experimented on a dataset combination of 70 COVID-19 images from one source~\cite{cohen2020covid} and non-COVID-19 images from Kaggle chest X-ray dataset. They proposed the Bayesian CNN model, which improves the detection rate from 85.7\% to 92.9\% along with the VGG16 model~\cite{shi2020review}. Similarly, ~\cite{narin2020automatic} proposed modification of three pre-trained models: ResNet50, InceptionV3 and Inception-ResNetV2 considering chest X-ray images. However, they have used only 100 images to conduct that experiment, and the dataset was the combination of 50 Kaggle's chest X-ray images (Pneumonia) as COVID-19 patients and 50 normal chest X-ray images as Non-COVID-19 patients. The experimental result showed that in terms of accuracy, ResNet50 (98\%) outperformed InceptionV3 (97\%) and InceptionResNetV2 (87\%), respectively.\\ Additionally,~\cite{zhang2020covid} presented a ResNet model in their work where they considered data imbalance as one of the primary concerns. They have used 70 COVID-19 patients and 1008 Non-COVID-19 pneumonia patients from different data sources. The evaluation result showed 96\% sensitivity, 70.7\% specificity and .952 of AUC for ResNet.\\
\cite{wang2020covid} introduced a deep CNN based model known as COVID-Net, which attained 83.5\% accuracy to detect COVID-19 patients from X-ray images. Their study comprises a dataset which contains 5941 images. The data set incorporates chest X-ray images of 1203 healthy people, 931 people with bacterial pneumonia, 660 patients with viral pneumonia, and 45 patients with COVID-19.\\
In general, most of the current research uses chest X-ray images, which combine different data sizes. While the deep learning methods showed promising results on chest X-ray images, it is difficult to conclude that the same deep learning models will do better on CT scan images. Apart from this, other deep learning models need to be tested to conclude that models such as ResNet50 or modified VGG16 are the best model so far for the initial screening of patients with COVID-19 symptoms.
\subsection{CT based screening}
There are a number of literature, which considered chest CT images to distinguish between COVID-19 and Non-COVID-19 patients\cite{chen2020deep,zhang2020covid,jin2020development,song2020deep,butt2020deep,li2020artificial}. ~\cite{chen2020deep} uses UNet++ to classify COVID-19 and Non-COVID-19 patients considering 132 sample images. The dataset includes 51 COVID-19, 55 Non-COVID-19, 16 viral pneumonia and 11 non-pneumonia patients. Their work also revealed that, using artificial intelligence it is possible to reduce the reading time of radiologists up to 65\%.
Another work conducted by~\cite{zhang2020covid} uses the UNet model for lung segmentation. They have used 540 images, which includes 313 with COVID-19 and 229 Non-COVID-19 patients’ images. In addition, a 3D CNN based model was proposed, which achieved 90.7\% of sensitivity and 91.1\% of specificity.\\
Beside 3D networks,~\cite{jin2020development} uses the ResNet15V2 model for detecting COVID-19 patients with the dataset of 1881 images (496 COVID-19 and 1385 Non-COVID-19). Their result acquired 94.1\% of sensitivity, while the specificity score was 95.5\%.
Several studies employed ResNet50 to detect COVID-19 patients from chest CT scan~\cite{song2020deep,butt2020deep}. For example,~\cite{song2020deep} considered 88 COVID-19, 100 bacterial pneumonia and 55 viral pneumonia images to classify between COVID-19 and other patients with ResNet50 on CT scan images. Using this technique, they have achieved 86\% accuracy. \cite{li2020artificial} uses ResNet-50 on a dataset containing 468 COVID-19 and 2996 other patients. Their experimental result showed, 90\% of sensitivity and 96\% of specificity on CT scan image dataset.\\
However, it is not surprising that, shortly, to handle many patients, a system may need to develop, which could screen COVID-19 patients using both CT scan and chest X-ray images.
\subsection{AI Explainable}
Explanatory Artificial intelligence (EAI) recently explored due to the ability to provide insights into the behavior and thought process of some sophisticated machine learning problems~\cite{ gilpin2018explaining}. Several studies showed, using a decision tree, linear models, it is possible to explain the models which are easily understandable and interpretable for humans~\cite{guidotti2018survey}.~\cite{ribeiro2016should} proposed Local Interpretable Model-agnostic Explanations (LIME). This novel explanation approach can explain the predictions of any classifier in an interpretable manner. They have explained the prediction of Google's pre-trained Inception neural network on arbitrary images in their work.\\ Using LIME, it is possible to understand, visualize, and interpret any deep learning models that used for image classification~\cite{samek2017explainable}.~\cite{holzinger2017we} demonstrated that explainable-AI helps to facilitate the implementation of AI/ML in the medical domain. Thus, an AI that explains the X-ray images' top features might play a crucial role in distinguishing between COVID-19 patients with other patients for the radiologist.\\
In general, this investigation found that a large portion of the study either considered chest X-ray or CT scan image analysis with a couple of deep learning models because of the time constraints. Be that as it may, when this experiment was conducted, none of the past works, considered both CT scan and chest X-ray images with different deep learning approaches to locate the best model on both scenarios with the help of AI-explanation. In this manner, this research expects to provide some insights on these issues, where eight diverse deep learning models were tried on both datasets. The experimental result will help the researchers and users develop a universal COVID-19 screening system using a deep learning application that could simultaneously work well on CT scan and chest X-ray images.
\section{Methodology}
Over the years, many studies use the Kaggle dataset as a reliable source for the experiment~\cite{narayanan2011link, iglovikov2017satellite, sutton2018nomad, mangal2016using, kumar2019sentiment}. Several deep learning approaches such as VGG16 and ResNet50 becomes popular when used for Kaggle dataset competition, and later those datasets were used for conducting various experiment~\cite{dhankhar2019resnet, zhaoclassification, savoiu2017recognizing}. In this pandemic situation, some of the literature relies on images (i.e., chest X-ray, CT scan) acquired from Kaggle datasets in order to develop an AI-based screening system for patients' with COVID-19 symptoms~\cite{tang2020rapidly, wang2020covid, sethy2020detection, narin2020automatic,apostolopoulos2020covid}. This research also used  images (chest X-ray and CT scan images) obtained from Kaggle datasets~\cite{Kaggle2020covid}. However, the dataset is the combination of multiple datasets. For example, 400 chest X-ray images containing both COVID-19 and Non-COVID-19 patients were collected from one source, and CT scan images containing 400 images were collected from another source. Finally, both datasets were examined separately with different deep learning techniques-VGG16~\cite{simonyan2014very}, InceptionResNetV2~\cite{langkvist2014review}, ResNet50~\cite{akiba2017extremely}, DenseNet201~\cite{huang2017densely}, VGG19~\cite{simonyan2014very}, MobileNetV2~\cite{sandler2018mobilenetv2}, NasNetMobile~\cite{da2018lung}, and ResNet15V2~\cite{varatharasan2019improving}.\\
Table~\ref{tab:table1} summarize two datasets used during this experiment. 80\% images were used for training, and 20\% were used for testing. Table 1 shows that, for each dataset, 320 images were used on the train set and 80 images were used for the test set.\\
\begin{table}[h]
 \caption{Dataset used during this experiment}
    \centering
    \begin{tabular}{cccc}\toprule
         Dataset&Label& Train Set& Test Set  \\\midrule
         \multirow{2}{*}{CT Scan}& COVID-19&160&40\\
         &Non-COVID-19&160&40\\\cline{2-4}
         \multirow{2}{*}{Chest X-ray}& COVID-19&160&40\\
         & Non-COVID-19&160&40\\
         \bottomrule
    \end{tabular}
   
    \label{tab:table1}
\end{table}
\subsection{Pre-trained Convet}
\begin{figure}[H]
     \centering
     \includegraphics[scale=.6]{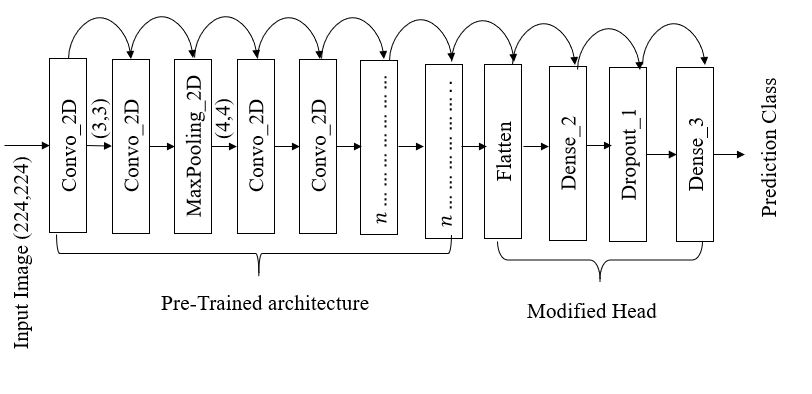}
     \caption{VGG16 architecture implemented during this experiment~\cite{simonyan2014very}}
     \label{fig:VGG16 architect}
 \end{figure}

\begin{figure}[h]
     \centering
     \includegraphics[width=\textwidth]{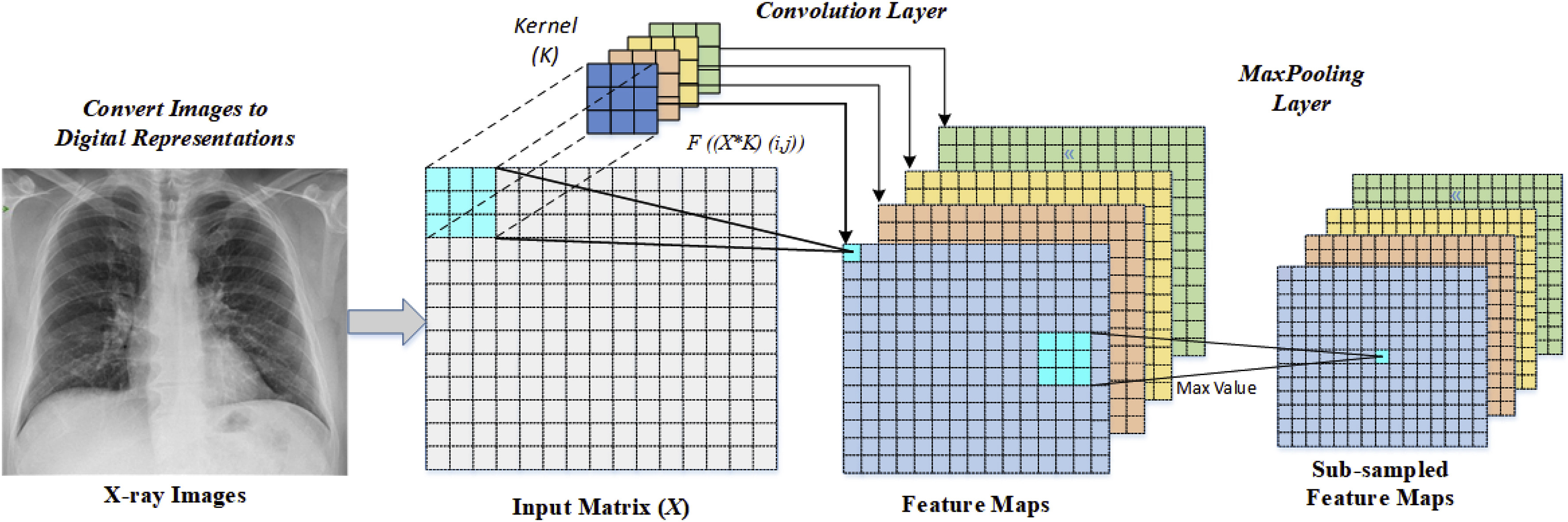}
     \caption{An illustration of convolutional and maxpooling layer operations~\cite{ozturk2020automated}}
     \label{fig:cnn_maxpool}
 \end{figure}

\begin{figure}[h]
    \centering
    \includegraphics[width=\textwidth]{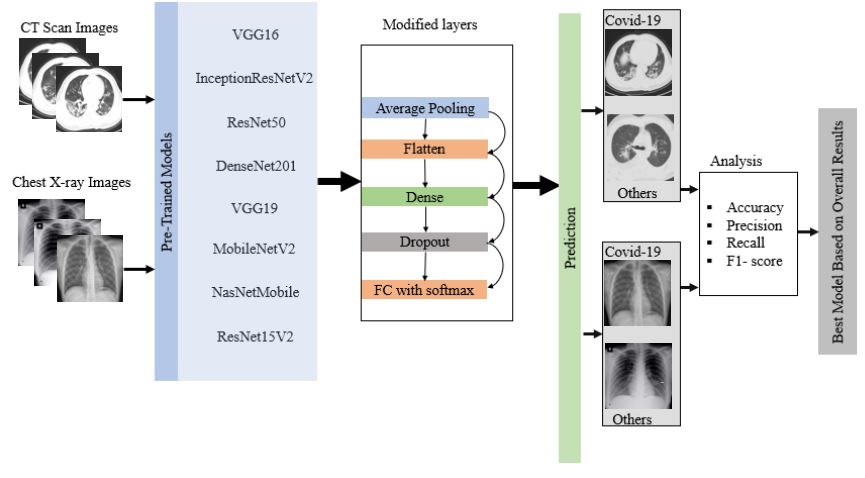}
    \caption{Flow diagram of the overall experiment}
    \label{fig:fig_1}
\end{figure}

Instead of developing a deep learning model from scratch, a more rational approach is to construct a model with existing, proven models~\cite{ozturk2020automated}. In this work we have uses the model’s weight from pre-trained models (i.e., imagenet~\cite{deng2009imagenet}) which developed as standard image recognition techniques~\cite{wu2019wider, abdel2013exploring, chennamsetty2018classification, chen2014convolutional}.
 This technique is also known as transfer learning, a process where a trained model on one problem could be used for other similar types of problems~\cite{pan2009survey}. This method's main advantage is less time consuming and possible to achieve higher accuracy with limited data~\cite{chollet2018deep}. When data sets are still developing and limited during this uncertain situation, like other studies~\cite{apostolopoulos2020covid, narin2020automatic, khalifa2020detection, sethy2020detection, abbas2020classification, chowdhury2020can}, this research takes into account that a pre-trained model would be the right choice. The primary model's architecture contains three components- pre-trained network, modified head, and prediction class (inspired from~\cite{zhang2020covid}). We employed the pre-trained network to extract its high-level features connected to the modified network and classification head, respectively.\\
 Most of the deep CNN models~\cite{ huang2017densely,sandler2018mobilenetv2,da2018lung} consists a number of convolution layers followed by the pooling layer (i.e., Maxpooling~\cite{graham2014fractional}, Averagepooling~\cite{ zubair2013dictionary}). Figure~\ref{fig:VGG16 architect}  illustrates the  modified architecture for VGG16. The architecture contains 16~\cite{yu2016visualizing} CNN layers with different filter numbers, sizes, and stride values.\\
 Let the letter,
 \begin{center} $c= convolutional layer$\\
 $m=maxpooling$\\
 $d=dense(fully connected layer)$\\\end{center}
 If we consider $c_1$ as the input layer, then our proposed models layout for VGG16 may express as:\\
 $c_1-c_2-m_1-c_3-c_4--m_n--c_n--d_1-d_2-d_3$
 A robust model also relay on proper feature extraction techniques as well\cite{gupta2017robust}.
 Let the letter,
 \begin{center}
 $ x=input\, image$\\
$k=kernel$\\\end{center}
Then the two-dimensional convolutional operation can be expressed as follows~\cite{ozturk2020automated}:
\begin{equation}
    (x*k)(i,j)=\sum_m\sum_nk(m,n)x(i-m,j-n)
\end{equation}
where * represents the discrete convolution operation~\cite{ozturk2020automated}. Kernel, K slides over the images with the stride parameters. The Rectified Linear unit (ReLu) is used as an activation function in the dense layer. ReLu function can be calculated with the following equations~\cite{ozturk2020automated}:
\begin{equation}
    f(x)=\begin{cases}
    0.01x  & \text{for $x<0$}\\
    x & \text{for $x\geq0$}\\
    \end{cases}
   \end{equation}
 During this experiment (3,3) convolution filter with (4,4) pool size is used for feature extraction~\cite{simonyan2014very}.
 An illustration for the flow of input image from convolutional layer and Maxpooling layer is given in~\ref{fig:cnn_maxpool}.
A deep learning model contains several tunable parameters (i.e., number of neurons, number of hidden layers)~\cite{denil2013predicting}. Since we have used a pre-trained model, thus batch size, the number of epochs and learning rate was only considered, instead of the number of neurons, and the number of hidden layers. However, manually tuning those parameters is time-consuming and less efficient~\cite{hutter2015beyond, qolomany2017parameters}. A better way to do it is to use grid search methods~\cite{bergstra2012random}. We have randomly selected the following parameters for our grid search methods:
\begin{center}
$Learning\, rate=[.001, .01, .1]$ \\
$Epochs=[10,20,30,40,50]$\\
$Batch\,size=[5,10,15,20]$ \\
\end{center}
Using the grid search method, we have found the following parameters as the most optimized parameters:
\begin{center}
 $Learning\, rate=.001$\\
 $Epochs=30$\\
 $Batch\, size=5$\\   
\end{center}
During the training phase, to optimize the model, we need to set an optimization algorithm~\cite{sutskever2013importance}. Some of the most popular optimization algorithms are- adaptive learning rate optimization algorithm (Adam)~\cite{kingma2014adam}, stochastic gradient descent (Sgd)~\cite{zhang2018theory}, and Root means square propagation (Rmsprop)~\cite{bengio2015rmsprop}. To keep our experiment simple, We have selected `Adam' as our optimization algorithm due to its effectiveness on binary image classification ~\cite{perez2017effectiveness, filipczuk2013computer}.\\
Finally, the overall result was statistically analyzed based on accuracy, precision, recall, and f1-score~\cite{ahsan2018real}\\
\begin{equation}\label{accuracy}
Accuracy=\frac{t_p+t_n}{t_p+t_n+f_p+f_n}\\
\end{equation}
\begin{equation}\label{precision}
 Precision=\frac{t_p}{t_p+f_p}\\   
\end{equation}
\begin{equation}\label{recall}
  Recall=\frac{t_p}{t_n+f_p}\\   
\end{equation}
\begin{equation}\label{f1-score}
F1=2\times\frac{Precision\times Recall}{Precision+Recall}
\end{equation}
Where,\\
True Positive ($t_p$)= COVID-19 patient classified as patient\\
False Positive ($f_p$)= Healthy people classified as patient\\
True Negative ($t_n$)=Healthy people classified as healthy\\
False Negative ($f_n$)= COVID-19 patient classified as healthy.\\
Figure~\ref{fig:fig_1} shows the overall flow diagram of the experiment. The best model was selected based on the statistical analysis on CT scan and chest X-ray image datasets.
\section{Result}
During this experiment overall accuracy, precision, recall, and f1-score were measured for eight different deep learning approaches considering CT scan and X-ray image using equation (\ref{accuracy}), (\ref{precision}), (\ref{recall}), and (\ref{f1-score}). We have used 80\% data for the train and 20\% data for the test, which is the most commonly used data mining techniques~\cite{mohanty2016using, menzies2006data, stolfo2000cost}. We ran the experiment twice and represented our result by averaging them for evaluation on the train and test set (inspired by~\cite{zhang2020covid}).
\subsection{CT Scan}
Table~\ref{tab:table_2} summarize the average accuracy, precision, recall, and f1-score for eight pre-trained deep learning models used during this experiment on the train set. The result shows that among all those model’s MobileNetV2 performed better in terms of accuracy (99\%), precision (99\%), recall (99\%), and f1-score (99\%) and ResNet50 demonstrated the worst performance- 56\% accuracy, 71\% precision, 56\% recall, and 47\% f1-score.
\begin{table}[H]
    \centering
    \begin{tabular}{ccccc}\toprule
         \multirow{2}{*}{Model}& \multicolumn{4}{c}{Performance}\\\cline{2-5}
         & Accuracy& Precision&Recall&F1-Score\\\midrule
         VGG16& 0.85&	0.85&	0.85&	0.85\\
         InceptionResNetV2& 0.81&	0.82&	0.81&	0.81\\
ResNet50&0.56&	0.71&	0.56&	0.47\\
DenseNet201&0.97&	0.97&	0.97&	0.97\\
VGG19&0.78&	0.82&	0.78&	0.77\\
MobileNetV2&0.99&	0.99&	0.99&	0.99\\
NasNetMobile&0.90&	0.90&	0.90&	0.90\\
ResNet15V2& 0.98&	0.98&	0.98&	0.98\\\bottomrule
    \end{tabular}
    \caption{Overall model’s performance on train set}
    \label{tab:table_2}
\end{table}
Table~\ref{tab:table_3} presents the overall model’s performance on a test set for CT scan images. The result manifests that, with 90\% accuracy, precision, recall, and f1-score NasNetMobile outperformed all other deep learning techniques. On the other hand, ResNet50 showed the worst performance with 55\% accuracy,64\% precision, 55\% of recall, and 46\% of f1-score.
\begin{table}[H]
    \centering
    \begin{tabular}{ccccc}\toprule
         \multirow{2}{*}{Model}& \multicolumn{4}{c}{Performance}\\\cline{2-5}
         & Accuracy& Precision&Recall&F1-Score\\\midrule
         VGG16& 0.86&	0.85&	0.86&	0.86\\
         InceptionResNetV2&0.84&	0.84&	0.84&	0.84\\
ResNet50&0.55&	0.64&	0.55&	0.46 \\
DenseNet201&0.79&	0.79&	0.79&	0.79\\
VGG19& 0.76&	0.81&	0.76&	0.75\\
MobileNetV2&0.89&	0.89&	0.89&	0.89\\
NasNetMobile&0.90&	0.90&	0.90&	0.90\\
ResNet15V2&0.84&	0.84&	0.84&	0.84\\\bottomrule
    \end{tabular}
    \caption{Overall model’s performance on test set}
    \label{tab:table_3}
\end{table}
\subsection{X-ray Image}
Table~\ref{tab:table_4} summarize the overall model’s execution amid this experiment on train sets for X-ray images. From the result- most of the model performed well on the train set except ResNet50. 100\% accuracy, precision, recall, and f1-score were achieved for VGG16, DenseNet201, MobileNetV2, NasNetMobile, and for ResNet15V2.
\begin{table}[H]
    \centering
    \begin{tabular}{ccccc}\toprule
         \multirow{2}{*}{Model}& \multicolumn{4}{c}{Performance}\\\cline{2-5}
         & Accuracy& Precision&Recall&F1-Score\\\midrule
         VGG16& 1.0&	1.0&	1.0&	1.0\\
         InceptionResNetV2& 0.99&	0.99&	0.99&	0.99\\
ResNet50& 0.64&	0.79&	0.64&	0.58\\
DenseNet201&1.0&	1.0&	1.0&	1.0\\
VGG19&0.98&	0.98&	0.98&	0.98\\
MobileNetV2& 1.0&	1.0&	1.0&	1.0\\
NasNetMobile&1.0&	1.0&	1.0&	1.0\\
ResNet15V2& 1.0&	1.0&	1.0&	1.0\\\bottomrule
    \end{tabular}
    \caption{Overall model's performance on train set}
    \label{tab:table_4}
\end{table}
Table~\ref{tab:table_5} summarize the model’s performance on the test set. Among all of the models, NasNetMobile appeared higher execution with 100\% accuracy, precision, recall and f1-score. On the other hand, ResNet50 appeared to have lower execution compared to any other model used during this test. The average accuracy, precision, recall, and f1-score for ResNet50 is 64\%, 79\%, 64\%, and 58\%, respectively.
\begin{table}[H]
    \centering
    \begin{tabular}{ccccc}\toprule
         \multirow{2}{*}{Model}& \multicolumn{4}{c}{Performance}\\\cline{2-5}
         & Accuracy& Precision&Recall&F1-Score\\\midrule
         VGG16& 0.97&	0.98&	0.97&	0.97\\
         InceptionResNetV2&0.97&	0.98&	0.97&	0.97\\
ResNet50&0.64&	0.79&	0.64&	0.58\\
DenseNet201&0.97&	0.98&	0.97&	0.97\\
VGG19&0.91&	0.93&	0.91&	0.91\\
MobileNetV2&0.97&	0.97&	0.97&	0.97\\
NasNetMobile&1.0&	1.0&	1.0&	1.0\\
ResNet15V2&0.99&	0.99&	0.99&	0.99\\\bottomrule
    \end{tabular}
    \caption{Overall model’s performance on test set}
    \label{tab:table_5}
\end{table}
\subsection{Confusion Matrix}
\begin{figure}[h]
    \centering
    \includegraphics[width=\textwidth]{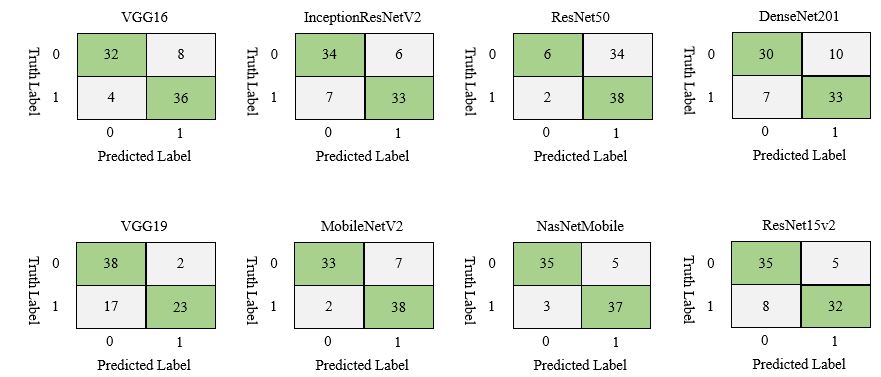}
    \caption{Confusion matrix of different deep learning model for CT scan image dataset}
    \label{fig:CT confusion matrix}
\end{figure}

The confusion matrix was calculated on the test set to simplify the understanding of the model's performance. Figure~\ref{fig:CT confusion matrix} depicts the confusion matrix for different models on given CT scan images. In the confusion matrix, the green box represented correctly classified images, and the white box represented the number of miss classification occurred by each model. Results show that NasNetMobile demonstrated the best result with only 8 misclassifications out of 80 images. On the other hand, with 36 misclassifications, ResNet50 shows the worst performance.
\\
Following figure~\ref{fig:chest x-ray confusion matrix} outlines the overall correctly and incorrectly classified chest X-ray images by eight deep learning models used during this work. Results show that most of the models performed well in chest X-ray images compared to CT scan images. Among all of the models, NasNetMobile was able to distinguish 100\% accurately both COVID-19 and other patients' X-ray images. Other algorithms, such as ResNet15V2 misclassified one image and VGG16, InceptionResNetV2, DenseNet201, and MobileNetV2 misclassified two images. Simultaneously, model ResNet50 and VGG19 showed poor performance and misclassified 29 and 7 images on chest X-ray images.
\begin{figure}[h]
    \centering
    \includegraphics[width=\textwidth]{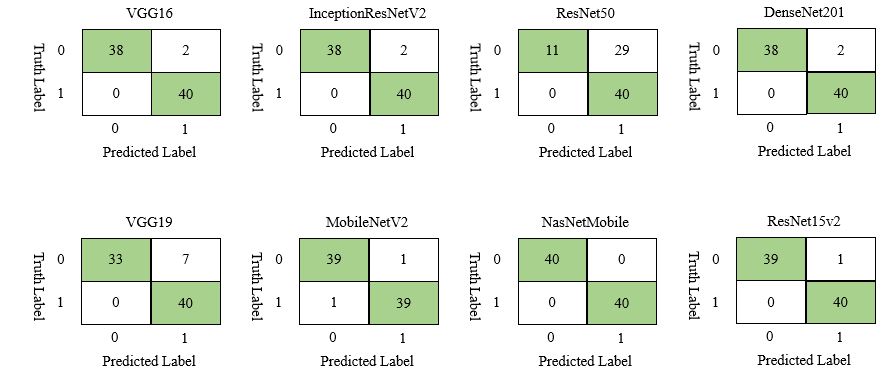}
    \caption{Confusion matrix of different deep learning model for chest x-ray image dataset}
    \label{fig:chest x-ray confusion matrix}
\end{figure}
\section{Models' Performance}
Each model's performance monitored on each epoch for both CT scan and chest X-ray images. The accuracy and loss were observed both on the train and test set.
\subsection{Considering CT scan image dataset}
\subsubsection{Models training and validation accuracy}
\begin{figure*}[]
    \centering
    \includegraphics[width=\textwidth]{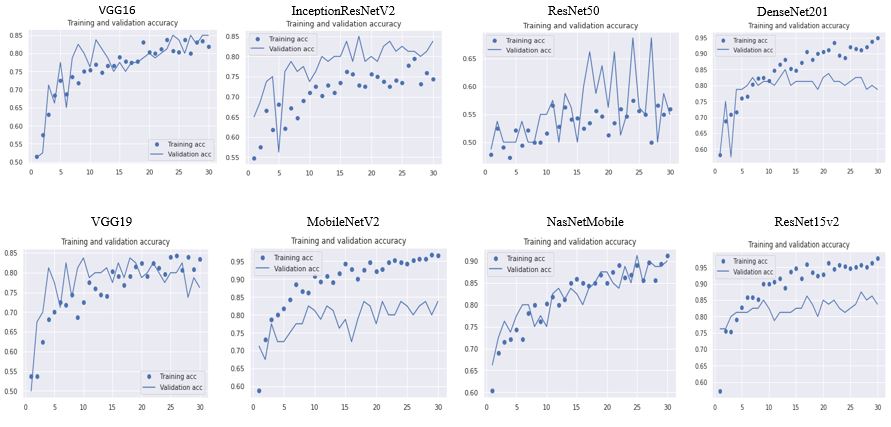}
    \caption{Models' accuracy during experiment for each epoch on CT scan image dataset}
    \label{models accuracy CT scan image}
\end{figure*}

Figure~\ref{models accuracy CT scan image} depicts the model's training and validation accuracy during each epoch. It shows that both training and validation follows the same trend for VGG16, InceptionResNetV2, VGG19, MobileNetV2, and NasNetMobile. However, on DenseNet201, ResNet15V2, after several epochs, training accuracy and validation accuracy started to disperse. For example, on ResNet15V2, while training accuracy reached up to 95\%, validation accuracy just fluctuated between 75\% to 80\% and started to drop below 80\%. Apart from this, for model ResNet50, both training and validation accuracy showed unsteadiness after epoch 15 to until epoch 30 and unable to achieve accuracy more than 65\% over time. 
\subsubsection{Models' training and validation loss}
\begin{figure}[h]
    \centering
    \includegraphics[width=\textwidth]{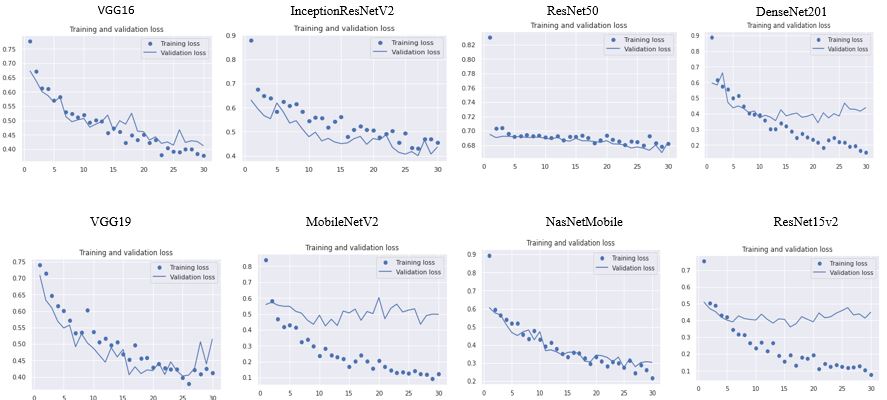}
    \caption{Models' loss during experiment for each epoch on CT scan image dataset}
    \label{fig:CT scan models' loss}
\end{figure}

Figure~\ref{fig:CT scan models' loss} delineates the model's training and validation loss for each epoch. It demonstrates that, both training and validation loss follows almost similar pattern for VGG16, InceptionResNetV2, ResNet50, VGG19, MobileNetV2, and NasNetMobile. Be that as it may, on DenseNet201, and ResNet15v2, validation, and training loss began to scatter after several epochs. For instance, on ResNet15V2, after 10 epochs, while training loss continuously decreased, the validation loss started to increase and rise to 50\%
\subsection{Considering chest X-ray image dataset}
\subsubsection{Models' training and validation accuracy}
\begin{figure}[h]
    \centering
    \includegraphics[width=\textwidth]{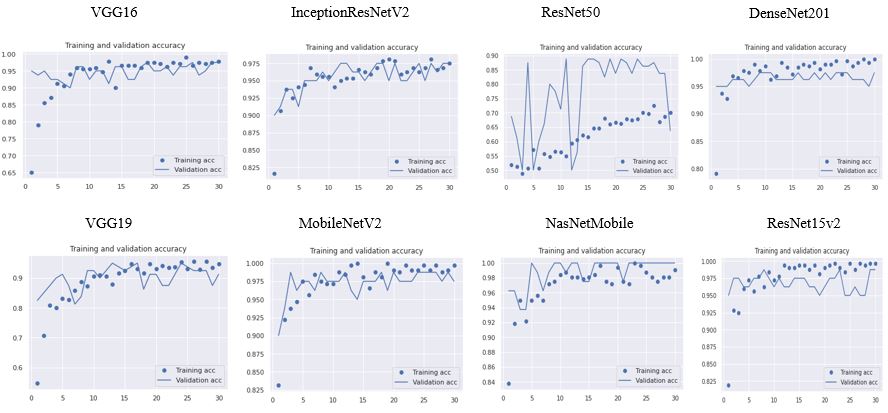}
    \caption{Models' accuracy during experiment for each epoch on chest X-ray image dataset}
    \label{fig:models' accuracy on chest x-ray}
\end{figure}

Figure~\ref{fig:models' accuracy on chest x-ray} shows the eight different deep learning model’s training and validation accuracy during each epoch. From figure~\ref{fig:models' accuracy on chest x-ray}, on model VGG16, InceptionResNetV2, and NasNetMobile, both the train set and test set follow the same trend. On the other hand, models like DenseNet201, VGG19, MobileNetV2, and ResNet15V2 also showed promising results even though models performance fluctuates after every five epochs. Among all of the models, ResNet50 showed the worst performance, and after 25 epoch, models training accuracy significantly dropped from 90\% to 60\% while validation accuracy was still rising. Based on the overall performance, it could be concluded that in terms of accuracy on train set and validation set, VGG16, InceptionResNetV2, and NasNetMobile showed more stability and better accuracy than the other six pre-trained convet.
\subsubsection{Model’s training and validation loss}
\begin{figure}[h]
    \centering
    \includegraphics[width=\textwidth]{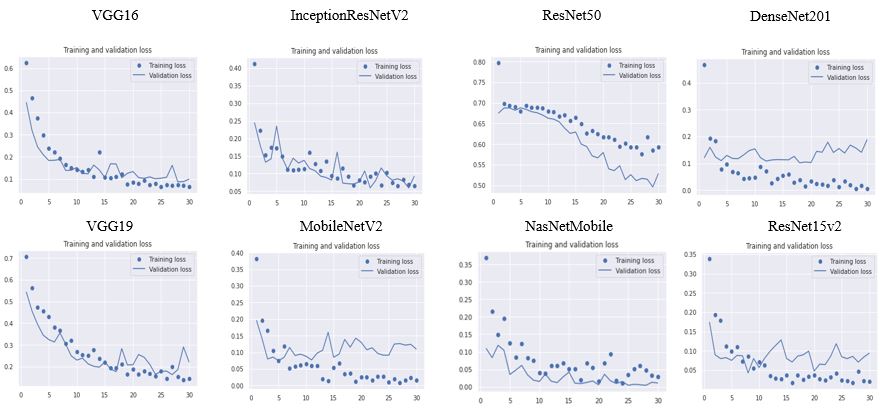}
    \caption{Models' loss during experiment for each epoch on chest X-ray image dataset}
    \label{fig:chest x-ray models' loss}
\end{figure}

Figure~\ref{fig:chest x-ray models' loss} demonstrates eight different deep learning models training and validation loss during each epoch.Here, model VGG16, InceptionResNetV2, VGG19, and NasNetMobile follow a similar pattern; contrary, ResNet50, DenseNet201, MobileNetV2, and ResNet15V2  demonstrates different patterns both on train and validation set. For example, on DenseNet201 after epoch 20, while train loss continuously dropped, the validation loss started to increase. Based on the overall experiment, it could be concluded that model VGG16, InceptionResNetV2, VGG19, and NasNetMobile showed better results and stability in terms of validation loss.
\subsection{Confidence Interval}
The Confidence interval (CI) was measured using two common methods, such as Wilson score~\cite{wilson1927probable} and Bayesian Interval~\cite{edwards1963bayesian}; both methods are widely used and showed better performance on the small dataset~\cite{brownlee2014machine}. Table~\ref{tab:CI} delineates 95\% (CI) for model accuracy on the test set for CT scan and chest x-ray images.
On the ct scan image dataset, Resnet50 has the lowest accuracy vary from 0.441 to 0.654 and o.441 to 0.656; in contrast, NasNetmobile has the highest accuracy set out from 0.815 to 0.948 and 0.820 to 0.952 respectively.\\
On the Chest X-ray image dataset, accuracy for VGG16, InceptionResNetV2, DenseNet201, and MobileNetV2 was achieved between 0.913 to 0.993, and 0.922 to 0.995 using Wilson score and Bayesian interval respectively. However, among all the models, Higher accuracy was obtained for NasNetMobile, and lower accuracy was acquired for ResNet50.   
\begin{table}[h!]
    \centering
    
    \caption{Confidence Interval ($\alpha = 0.05$) of CT scan and Chest X-ray in terms of accuracy on test set. Sample size, n=80 for both study.}
    \vspace*{\baselineskip}
    \begin{tabular}{ccccc}\toprule
        \multirow{3}{*}{Study} &\multirow{3}{*}{Model}& Test& \multicolumn{2}{c}{\multirow{2}{*}{\textit{Methods}}} \\
         &&accuracy\\\cmidrule{4-5}
         &&& Wilson Score& Bayesian Interval\\ \midrule
         \multirow{6}{*}{CT scan}& VGG16&0.86&0.756 -- 0.912&0.76 -- 0.915\\
         &InceptionResNetV2&0.84&0.742 -- 0.903& 0.745 -- 0.906 \\
         &ResNet50&0.55&0.441 -- 0.654& 0.441 -- 0.656 \\
         &DenseNet201&0.79&0.686 -- 0.863& 0.689 -- 0.866\\
         &VGG19&0.76& 0.659 -- 0.842& 0.661 -- 0.845\\
         &MobileNetV2&0.89& 0.800 -- 0.940& 0.805 -- 0.943\\
         &NasNetMobile&0.90&0.815 -- 0.948& 0.820 -- 0.952\\
         &ResNet15V2&0.84& 0.742 -- 0.903& 0.745 -- 0.906\\\bottomrule
         \multirow{6}{*}{Chest X-ray}& VGG16&0.97&0.913 -- 0.993& 0.922 -- 0.995\\
         &InceptionResNetV2&0.97&0.913 -- 0.993& 0.922 -- 0.995\\
         &ResNet50&0.64&0.528 -- 0.734& 0.529 -- 0.736 \\
         &DenseNet201&0.97&0.913 -- 0.993& 0.922 -- 0.995\\
         &MobileNetV2&0.97&0.913 -- 0.993& 0.922 -- 0.995\\
         &NasNetMobile&1.0& 0.954 -- 1.00& 0.969 -- 1.00\\
         &ResNet15V2&0.99&0.933 -- 0.998& 0.943 -- 0.999\\\bottomrule
    \end{tabular}
    
    \label{tab:CI}
\end{table}\\

\section{Discussion}
During this experiment, extensive analyses were done, considering both CT scan images and chest X-ray images using eight different deep learning approaches. One of this section's main goals is to find out the best model considering both CT scan and chest X-ray images.
\subsection{Best Model on CT Scan Image Dataset}
The best models for this specific experiment were selected considering the following factors- accuracy, precision, recall, f1 score, train and validation loss, and performance of confusion matrix.
On the CT scan image dataset, on the train set, MobileNetV2 showed higher accuracy, precision, recall, and f1-score as 99\%. However, on the test set, NasNetMobile outperformed all other models with 90\% accuracy, precision, recall, and f1-score.\\
On the other hand, when we looked over the confusion matrix result, it was found that, compared to any model, NasNetMobile demonstrated better results with only 8 misclassifications out of 80 images. However, model MobielNetV2 misclassified 13 of the images out of 80 images. Additionally, VGG16, InceptionResNetV2, VGG19, MobileNetV2, and NasNetMobile performed better in training and validation accuracy than any other model. Contrary, in terms of training and validation loss, VGG16, InceptionResNetV2, ResNet50, VGG19, MobileNetV2, and NasNetMobile showed a better result.
From table~\ref{tab:overallsummaryonctimage}, we can see that, compared to any model, MobileNetV2 outperformed all other models in terms of accuracy, precision, recall, and f1-score on train set and NasNetMobile on the test set. However, considering the confusion matrix, NasNetMobile outperformed all other methods. Additionally, the misclassification difference between MobileNetV2 and NasNetMobile is just one. Thus, it could be concluded that considering all those factors, both MobileNetV2 and NasNetMobile are the best models on the CT scan image dataset.
\begin{table}[h]

    \centering
    \resizebox{\textwidth}{!}{
    \begin{tabular}{@{}ccccccccc@{}c@{}cc@{}}\toprule
         \multirow{2}{*}{Model}&\multicolumn{2}{c}{Accuracy}&\multicolumn{2}{c}{Precision}&\multicolumn{2}{c}{Recall}&\multicolumn{2}{c}{F1-Score}&{Confusion Matrix}&\multicolumn{2}{c}{\makecell{Accuracy and Loss\\ During Epochs}}  \\\cmidrule{2-9}\cmidrule{11-12}
         & Train&Test&Train&Test&Train&Test&Train&Test&Misclassified& Accuracy& Loss\\\midrule
         VGG16&85\%	&86\%&	85\%&	85\%&	85\%&	86\%&	85\%&	86\%&	12&	Satisfactory&	Satisfactory\\
InceptionResNetV2& 81\%&	84\%&	82\%&	84\%&	81\%&	84\%&	81\%&	84\%&	13&	Satisfactory&	Satisfactory\\
ResNet50&56\%&	55\%&	71\%&	64\%&	56\%&	55\%&	47\%&	46\%&	36&	Not satisfactory&	Satisfactory\\
VGG19&78\%&	76\%&	82\%&	81\%&	78\%&	76\%&	77\%&	75\%&	19&	Satisfactory&	Satisfactory\\
MobileNetV2&99\%&	89\%&	99\%&	89\%&	99\%&	89\%&	99\%&	89\%&	9&	Satisfactory&	Satisfactory\\
NasNetMobile&90\%&	90\%&	90\%&	90\%&	90\%&	90\%&	90\%&	90\%&	8&	Satisfactory&	Satisfactory\\\bottomrule
    \end{tabular}}
    \caption{Overall summary of the best model found considering various factor on CT scan image dataset}
    \label{tab:overallsummaryonctimage}
\end{table}
\subsection{Best Model on X-ray Image Dataset}
The best model on X-ray image dataset was chosen by following the same procedure as the CT scan image dataset.\\
Based on the train set's overall performance, VGG16, DenseNet201, MobileNetV2, NasNetMobile, and ResNet15V2 outperformed all other models in terms of accuracy, precision, recall, and f1-score. On the other hand, on the test set, NasNetMobile outperformed all other models. Apart from this, based on the confusion matrix result, NasNetMobile exceeded all other models with zero misclassification.\\
However, while considering the model’s training and validation accuracy during each epoch, it was seen that VGG16, InceptionResNetV2, and NasNetMobile outperformed other models. Additionally, during training and validation loss, VGG16, InceptionResNetV2, VGG19, and NasNetMobile showed better results than other models.
To find out the best model, a comparison was made on the following table~\ref{tab:overall summary on chest x-ray image}.
Table~\ref{tab:overall summary on chest x-ray image} showed that NasNetMobile outperformed all of the models taking into account all performance measurement tools such as accuracy (100\%), precision (100\%), recall (100\%), f1-score (100\%), confusing matrix (100\%), and loss calculation.\\
\begin{table}[h]
    \centering
    \resizebox{\textwidth}{!}{
    \begin{tabular}{cccccccccccc}\toprule
         \multirow{2}{*}{Model}&\multicolumn{2}{c}{Accuracy}&\multicolumn{2}{c}{Precision}&\multicolumn{2}{c}{Recall}&\multicolumn{2}{c}{F1-Score}&{Confusion Matrix}&\multicolumn{2}{c}{\makecell{Accuracy and Loss\\ During  Epochs}}  \\\cmidrule{2-9}\cmidrule{11-12}
         & Train&Test&Train&Test&Train&Test&Train&Test&Misclassified& Accuracy& Loss\\\midrule
         MobileNetV2&	100\%&	97\%&	100\%&	97\%&	100\%&	97\%&	100\%&	97\%&	2&	Not satisfactory&	Not satisfactory\\
         ResNet15V2&	100\%&	99\%&	100\%&	99\%&	100\%&	99\%&	100\%&	99\%&	1&	Not satisfactory&	Not satisfactory\\
         DenseNet201&	100\%&	97\%&	100\%&	98\%&	100\%&	97\%&	100\%&	97\%&	2&	Not satisfactory&	Not satisfactory\\
         VGG16&	98\%&	97\%&	98\%&	98\%&	98\%&	97\%&	98\%&	97\%&	2&	Satisfactory&	Satisfactory\\
         InceptionResNetV2&	99\%&	97\%&	99\%&	98\%&	99\%&	97\%&	99\%&	97\%&	2&	Satisfactory&	Satisfactory\\
         NasNetMobile&	100\%&	100\%&	100\%&	100\%&	100\%&	100\%&	100\%&	100\%&	0&	Satisfactory&	Satisfactory\\
         VGG19&	98\%&	91\%&	98\%&	93\%&	98\%&	91\%&	98\%&	91\%&	7&	Not satisfactory&	Satisfactory\\\bottomrule
    \end{tabular}}
    \caption{Overall summary of the best model found considering various factor on chest X-ray image dataset}
    \label{tab:overall summary on chest x-ray image}
\end{table}\\
Based on all the models' overall performance, it was found that while model MobileNetV2 showed best results on the CT scan image dataset, model NasNetMobile showed better results on the Chest x-ray image dataset. To find out the best model among two of those, a comparison was made between those two models. The models' performance was calculated by averaging the overall performance on both the train set and the test set. For example, the average accuracy of Model NasNetMobile on datasets was calculated as follows:\\
\begin{center}
 $Accuracy=\frac{\makecell{ X\-ray\,accuracy (train+test)\\+CT scan\,accuracy(train+test)}}{4}$\\
 \vspace{5pt}
$=\frac{100\%+100\%+90\%+90\%}{4}$
$=95\%$
\end{center}
From table~\ref{tab:comparision between mobilenvt vs nasnet}, we can see that MobileNetV2 outperformed NasNetMobile in terms of accuracy, precision, recall, and f1-score. However, misclassification rate for MobileNetV2 (11.25\%) is slightly higher than NasNetMobile (10\%). Since our dataset is small, this error rate may not be significant, yet, for a larger dataset, the misclassification rate may significantly impact.
\begin{table}[H]
\centering
    \begin{tabular}{cccccc}\toprule
         Model&	Accuracy&	Precision&	Recall&	F1-score&	\makecell{Error rate\\( test set) } \\\midrule
         MobileNetV2&	96.25\%&	96.25\%&	96.25\%&	96.25\%&	11.25\%\\
         NasNetMobile&	95\%&	95\%&	95\%&	95\%&	10\%\\\bottomrule
    \end{tabular}
    \caption{Comparison between MobileNetV2 and NasNetMobile on both dataset}
    \label{tab:comparision between mobilenvt vs nasnet}
\end{table}
\subsection{Models Average Accuracy}
Average accuracy was calculated by averaging the training and testing accuracy of all the models. Table~\ref{tab:models average accuracy} shows the average accuracy for CT scan and chest X-ray image dataset. Results show that almost all models performed better on the X-ray image data set compared to the CT scan. The average accuracy for all the models on CT scan and X-ray image dataset is 82.94\% and 93.94\%, respectively.
\begin{table}[H]
    \centering
    \begin{tabular}{ccc}\toprule
         Model&	CT\, scan&	X-ray  \\\midrule
         VGG16&	$\frac{(85+86)}{2}= 85.5\%$&	$\frac{(100+97)}{2}= 98.5\%$\\
         InceptionResNetV2&	$\frac{(81+84)}{2}=82.5\%$&	$\frac{(99+97)}{2}=98\%$\\
    ResNet50&	$\frac{(56+55)}{2}=55.5\%$&	$\frac{(64+64)}{2}=64\%$\\ 
    DenseNet201&	$\frac{(97+79)}{2}=88\%$&$	\frac{(100+97)}{2}=98.5\%$\\
    VGG19&$\frac{(78+76)}{2}=77\%$&$\frac{(98+91)}{2}=94.5\%$\\
MobileNetV2&$\frac{(99+89)}{2}=94\%$ &$\frac{(100+97)}{2}=98.5\%$\\
NasNetMobile&$\frac{(90+90)}{2}=90\%$&$\frac{(100+100)}{2}=100\%$\\
ResNet15V2&$\frac{(98+84)}{2}=91\%$&$\frac{100+99}{2}=99.5\%$\\
Average&$82.94\%$&$93.94\%$\\\bottomrule
    \end{tabular}
    \caption{Models average accuracy on both dataset}
    \label{tab:models average accuracy}
\end{table}
\subsection{Feature Territory Highlighted by the Model on Different Layer}
\begin{figure}[h]
    \centering
    \includegraphics[width=\textwidth]{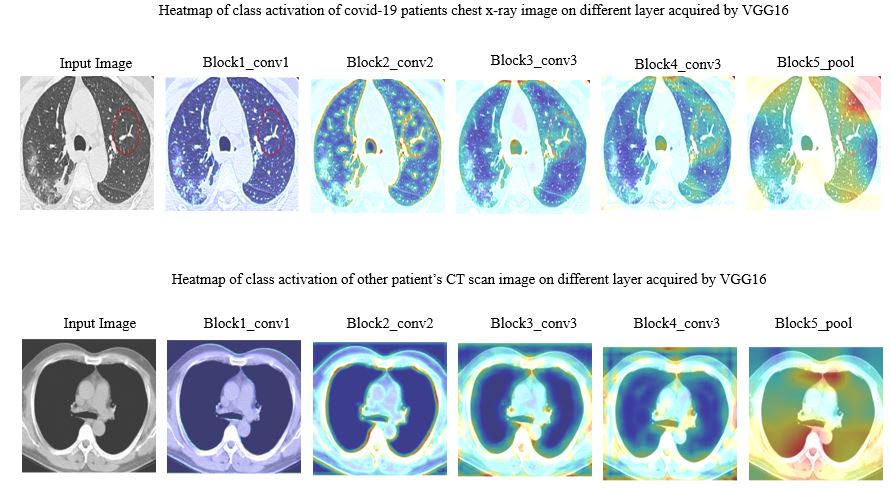}
    \caption{Heat map of class activation of CT scan image on different layer acquired by VGG16}
    \label{fig:Heatmap on CT scan}
\end{figure}

\begin{figure}[h]
    \centering
    \includegraphics[width=\textwidth]{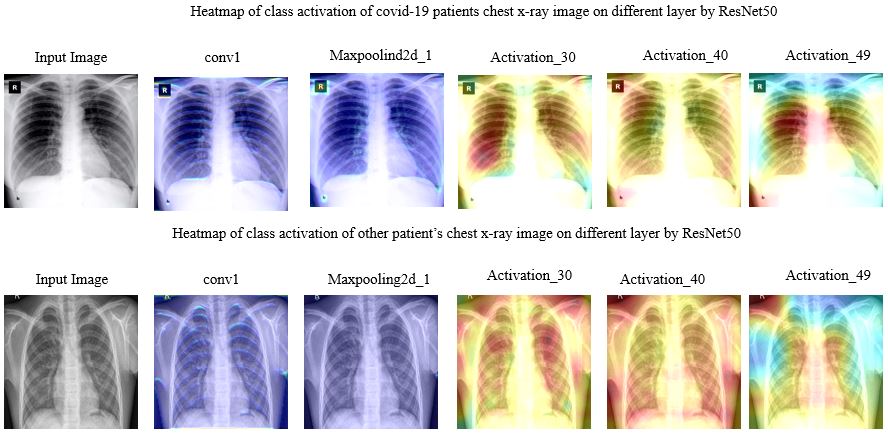}
    \caption{Heat map of class activation of chest X-ray  image on different layer acquired by ResNet50}
    \label{fig:heatmap chest xray}
\end{figure}

In this work, we tried to understand how each layer dealt with the actual image. Figure~\ref{fig:Heatmap on CT scan} demonstrated CT scan images during different layers. Note that, just a few of the layers from VGG16 were addressed here to give some insights.
Here another Figure~\ref{fig:heatmap chest xray}, manifested the different layer’s activity of model ResNet50 on chest X-ray images. The region spotted by model ResNet50 was highlighted with heatmap.
\subsection{Models Interpretability with LIME}
\begin{figure}[H]
    \centering
    \includegraphics[scale=.5]{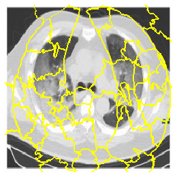}
    \caption{: Super-pixels on a sample chest CT scan images}
    \label{fig:super pixel CT scan}
\end{figure}

\begin{figure}[H]
    \centering
    \includegraphics[scale=.7]{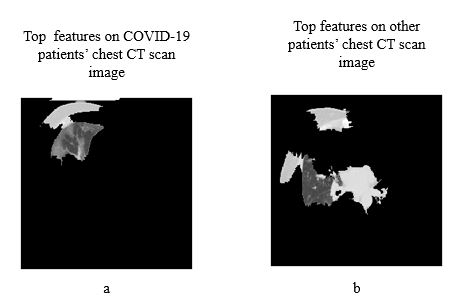}
    \caption{Top Four features a) on COVID-19 patients CT scan image b) on other patients CT scan image}
    \label{fig:CT top features}
\end{figure}

\begin{figure}[h]
    \centering
    \includegraphics[scale=.6]{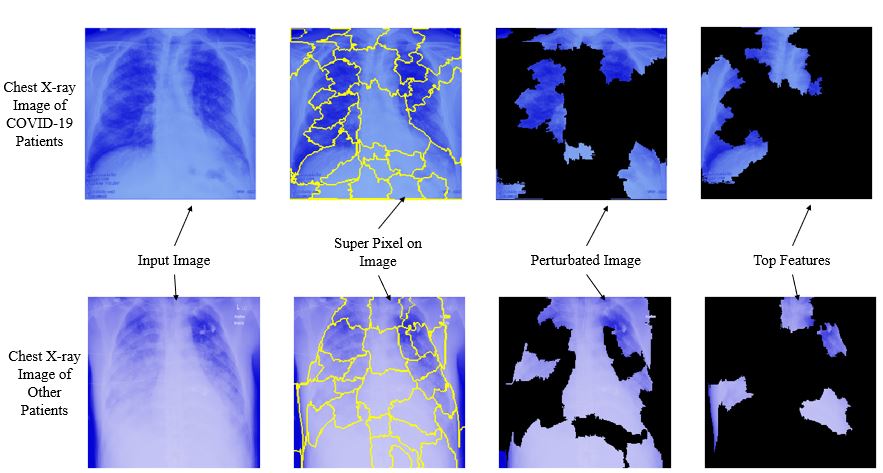}
    \caption{Overall prediction analysis using LIME}
    \label{fig:Chest X-ray with Lime}
\end{figure}

The pre-trained CNN model extracts sophisticated features from the images, which sometimes reveals unnecessary features. Therefor training such a model is often computationally expensive.
Additionally, a large set of features makes it challenging to understand which features are essential for predictions. In this paper, these issues also addressed while developing a model for screening COVID-19 patients.\\ 
To identify which specific features help deep learning model (MobileNetV2, NasNetMobile) to differentiate between COVID-19 and Non-COVID-19 patients, LIME was used. Local Interpretable Model-agnostic Explanations (LIME) is a procedure that helps to understand how the input features of a deep learning model affect its predictions. For example, for image classification, LIME finds the set of super-pixels with the most grounded relationship with a prediction label~\cite{khan2020cidmp}. LIME makes clarifications by creating another dataset of random perturbations (with their separate forecasts) around the occasion being clarified and afterward fitting a weighted neighborhood proxy model. This neighborhood model is usually a more straightforward model with natural interpretability, such as a linear regression model. LIME creates perturbations by turning on and off a portion of the super-pixels in the image. A quick shift strategy was utilized with the following parameters in order to calculate the super pixel, as shown in table~\ref{tab:super pixel}:
\begin{table}[H]
    \centering
    \begin{tabular}{cc}\toprule
         Function& value  \\\midrule
         Kernel Size& 4\\
         Maximum Distance& 200\\
         Ratio&0.2\\\bottomrule
    \end{tabular}
    \caption{Parameter used to calculate maximum pixel}
    \label{tab:super pixel}
\end{table}
Figure~\ref{fig:super pixel CT scan} is the output after computing the super-pixels on a sample chest CT scan images.
Additionally, the following figure~\ref{fig:CT perturbation} shows different image conditions considering perturbation vectors and perturbed images. To predict the class, during this experiment 150 perturbations were used.
\begin{figure}[H]
    \centering
    \includegraphics[scale=.5]{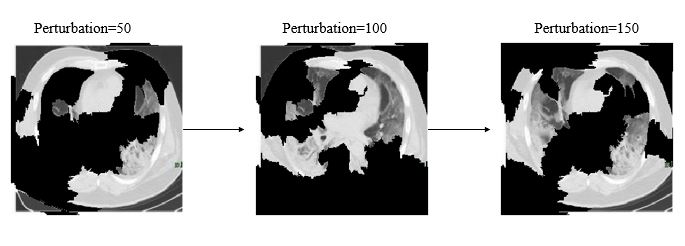}
    \caption{Examples of perturbation vectors and perturbated images}
    \label{fig:CT perturbation}
\end{figure}
The distance metric was utilized to assess how far each perturbation is from the original image. Cosine metrics were used with kernel width as 1/4 to measure the original image's distance and perturbed images. A weighted linear model was used to explain the overall model. A coefficient was found for every superpixel in the picture that represents how solid the superpixels impact in the prediction of COVID-19 patients. 
Finally, top features were sorted in order to determine what are the most important superpixel. Here, figure~\ref{fig:CT top features} demonstrates the top four critical features.
Here figure~\ref{fig:Chest X-ray with Lime}, depicts the overall interpretability for image classification with LIME on chest X-ray images considering each step. The prediction was conducted using NasNetMobile. Using LIME, it was possible to identify which top features helps to identify COVID-19 patients from other patients considering chest X-ray images.
\\
In brief, based on the overall experiment, this study found that, among all eight deep learning models, MobileNetv2 and NasNetMobile performed better both on CT scan and chest X-ray image datasets. Additionally, all deep learning models performed well on the chest X-ray image dataset compared to CT scan images with an average 8\% higher accuracy. This research addressed that existing deep learning approaches could be an alternative solution for detecting COVID-19 patients. However, a proper screening system should be developed based on the Expert like Doctors’ and Radiologists opinion as well.
\section{Conclusion}
During this pandemic situation, there are many countries and places where it is difficult to test enough patients with existing tool kit or CT scan images due to the expense and time sensitivity issues. Thus, as a helping hand, a comprehensive study was conducted considering CT scan (400 samples) and chest X-ray (400 samples) image dataset to classify between patients with COVID-19 symptoms with other patients. Additionally, top features that differentiate between COVID-19 and other patients were analyzed using LIME.  The experimental result revealed that existing deep-learning models performed better on chest X-ray images compared to CT scan images. Moreover, a chest X-ray takes less time and also turns out to be cost-efficient. Thus, a chest X-ray could be an alternative approach in order to resolve this shortage. Additionally, in this study, eight different deep learning models (VGG16, InceptionResNetV2, ResNet50, DenseNet201, VGG19, MobileNetV2, ResNet15V2, and NasNetMobile) were used and analyzed. The research outcome showed that, among all of the models, in CT scan image dataset, MobileNetV2 and NasNetMobile outperformed all other models, and NasNetMobile is the best model on chest X-ray image dataset. 
However, since the dataset is comparatively small, thus the accuracy acquired from the model may not represent the exact accuracy on a large scale. Therefore, 95\% CI for the accuracy on the test set was measured for all the models and results showed that NasNetMobile outperformed all other models with 95\% CI on CT scan datasets, accuracy ranges from 81.5\% to 95.2\%, and on chest X-ray image dataset varies from 95.4\% to 100\%.
The experimental result may bolster other current studies, which proposed that it is possible to develop an initial COVID-19 screening system using a deep-learning approach. With this short time and pandemic situations, we hope our study will give some insights to researchers and developers who are actively looking for alternative screening procedures by using both CT scan and chest X-ray image datasets for COVID-19 patients. Further study includes but not limited to- understanding deep learning models performance with highly imbalanced data, model performance with a larger dataset, Check for data bias\cite{senempirical}, parameter tuning, and developing a decision support system. 

\bibliographystyle{unsrt}  
\bibliography{main.bib}

\end{document}